\documentclass{PoS}

\usepackage{amssymb,amsmath,epsfig}

\renewcommand{\d}{\mathrm{d}}
\newcommand{\bm}[1]{{\bf #1}}

\newcommand{\mc}[1]{\mathcal{#1}}
\renewcommand{\sl}{\slashed}

\renewcommand{\d}{\mathrm{d}}
\newcommand{\sT}{{\scriptscriptstyle T}}
\newcommand{\pT}{\bm{p}_\sT}
\newcommand{\kT}{\bm{k}_\sT}
\newcommand{\qT}{\bm{q}_\sT}

\title{Determining the Higgs spin and parity in the di-photon decay channel 
using gluon polarization}

\ShortTitle{Determining the Higgs spin and parity using gluon polarization}

\author{\speaker{Wilco J. DEN DUNNEN}%
        \\
       ( Universit\"at T\"ubingen)\\
       E-mail: \email{wilco.den-dunnen@uni-tuebingen.de}}

\author{Marc SCHLEGEL\\
        ( Universit\"at T\"ubingen)\\
        E-mail: \email{marc.schlegel@uni-tuebingen.de}}

\abstract{
Gluons inside an unpolarized proton are in general linearly polarized.
This polarization can, in principle, be used to determine the spin and parity of the newly 
found Higgs-like boson at the LHC.
We focus here on the calculation of the degree of polarization.
}

\FullConference{XXI International Workshop on Deep-Inelastic Scattering and Related Subjects\\
		 22-26 April, 2013\\
		 Marseilles, France}

\begin{document}

With the discovery of a new Higgs-like particle at the LHC \cite{:2012gk,:2012gu}
the task at hand is to verify, as accurately as possible, whether all its 
properties are those predicted by the Standard Model (SM).
A specific property of the SM Higgs boson is its parity-even scalar coupling
to other particles.
This feature needs verification in all its production and decay channels independently.

In the gluon fusion production channel with a decay to two photons,
the full process is characterized by only one single angle in the absence of gluon
polarization.
This polar angle $\theta$, between the diphoton axis and digluon axis
in the partonic center of mass frame, 
does not contain any information on the parity of the coupling, 
nor can it be used to distinguish between all possible spin-2 coupling scenarios
\cite{Choi:2002jk,Gao:2010qx,Bolognesi:2012mm,Choi:2012yg,Ellis:2012jv}.
In the case that one or both of the incoming gluons are linearly polarized,
the azimuthal distribution can also be non-trivial and,
at the same time, will the gluon polarization modify the transverse momentum 
distribution of the produced Higgs-like particle. 

A gluon extracted from an unpolarized proton is in general linearly polarized
with a magnitude and direction depending on its transverse momentum. 
This effect is not taken into account in event generators,
where the partonic transverse momentum is generated by parton showers that
leave the gluons unpolarized.
In earlier publications we reported on the effect of this gluon polarization
on the transverse momentum distribution and azimuthal distribution of decay
products in Higgs production \cite{Boer:2011kf,denDunnen:2012ym,Boer:2013fca}
and how it can be used to determine its spin and parity.
Here we will focus on the calculation of the degree of polarization.

\section*{Transverse Momentum Dependent factorization}

To accurately describe Higgs boson production at small and moderate transverse
momentum one needs to use the framework of Transverse Momentum Dependent (TMD)
factorization.
In that framework, the full $pp\to \gamma\gamma X$ cross section is split into a partonic
$gg\to \gamma\gamma$ cross section and two TMD gluon correlators, which describe the 
distribution of gluons inside a proton as a function of not only its momentum along the
direction of the proton, but also transverse to it.
More specifically, the differential cross section for the inclusive 
production of a photon pair from gluon-gluon fusion is written as 
\cite{Ji:2005nu,Sun:2011iw,Ma:2012hh},
\begin{equation}\label{eq:factformula}
\frac{\d\sigma}{\d^4 q \d \Omega}
  \propto 
  \int\!\! \d^{2}\pT \d^{2}\kT
  \delta^{2}(\pT + \kT - \qT)
  \mc{M}_{\mu\rho \kappa\lambda}
  \left(\mc{M}_{\nu\sigma}^{\quad\kappa\lambda}\right)^*
  \\
  \Phi_g^{\mu\nu}(x_1,\pT,\zeta_1,\mu)\,
  \Phi_g^{\rho\sigma}(x_2,\kT,\zeta_2,\mu),
\end{equation}
with the longitudinal momentum fractions $x_1={q\cdot P_2}/{P_1\cdot P_2}$
and $x_2={q\cdot P_1}/{P_1\cdot P_2}$, $q$ the momentum of the photon pair, 
$\mc{M}$ the $gg\to \gamma\gamma$ partonic hard scattering matrix element
and $\Phi$ the following gluon TMD correlator in an unpolarized proton,
\begin{align}\label{eq:TMDcorrelator}
\Phi_g^{\mu\nu}(x,\pT,\zeta,\mu) 
&\equiv
	      \int \frac{\d(\xi\cdot P)\, \d^2 \xi_\sT}{(x P\cdot n)^2 (2\pi)^3}
	      e^{i ( xP + p_\sT) \cdot \xi}\,
	      \langle P| F_a^{n\nu}(0)
	      \left(\mc{U}_{[0,\xi]}^{n[\text{--}]}\right)_{ab} F_b^{n\mu}(\xi)
	      |P\rangle \Big|_{\xi \cdot P^\prime = 0}\nonumber\\
&=	-\frac{1}{2x} \bigg \{g_\sT^{\mu\nu} f_1^g(x,\pT^2,\zeta,\mu)
	-\bigg(\frac{p_\sT^\mu p_\sT^\nu}{M_p^2}\,
	{+}\,g_\sT^{\mu\nu}\frac{\pT^2}{2M_p^2}\bigg)
	h_1^{\perp\,g}(x,\pT^2,\zeta,\mu) \bigg \} + \text{HT},
\end{align}
with $p_{\sT}^2 = -\pT^2$ and $g^{\mu\nu}_{\sT} = g^{\mu\nu}
- P^{\mu}P^{\prime\nu}/P{\cdot}P^\prime - P^{\prime\mu}P^{\nu}/P{\cdot}P^\prime$,
where $P$ and $P^\prime$ are the momenta of the colliding protons and $M_p$ their mass.
The gauge link $\mc{U}_{[0,\xi]}^{n[\text{--}]}$ for this process arises 
from initial state interactions. 
It runs from $0$ to $\xi$ via minus infinity along the direction $n$, 
which is a time-like dimensionless four-vector with no transverse 
components such that $\zeta^2 = (2n{\cdot}P)^2/n^2$. 
In principle, Eqs.\ \eqref{eq:factformula} and \eqref{eq:TMDcorrelator} also contain
soft factors, but with the appropriate choice of $\zeta$ (of around 1.5 times $\sqrt{s}$), 
one can neglect their contribution, at least up to next-to-leading order 
\cite{Ji:2005nu,Ma:2012hh}.
To avoid the appearance of large logarithms in ${\cal M}$, 
the renormalization scale $\mu$ needs to be of order $M_h$.
The second line of Eq.\ \eqref{eq:TMDcorrelator} contains the parameterization of the
TMD correlator in terms of the unpolarized gluon distribution $f_1^g$,
the linearly polarized gluon distribution $h_1^{\perp\,g}$ and
Higher Twist (HT) terms, which only give $\mc{O}(1/Q)$ suppressed contributions to the
cross section, where $Q\equiv \sqrt{q^2}$.

\section*{Degree of Polarization}
\begin{figure}
\centering
\includegraphics[width=4.5cm]{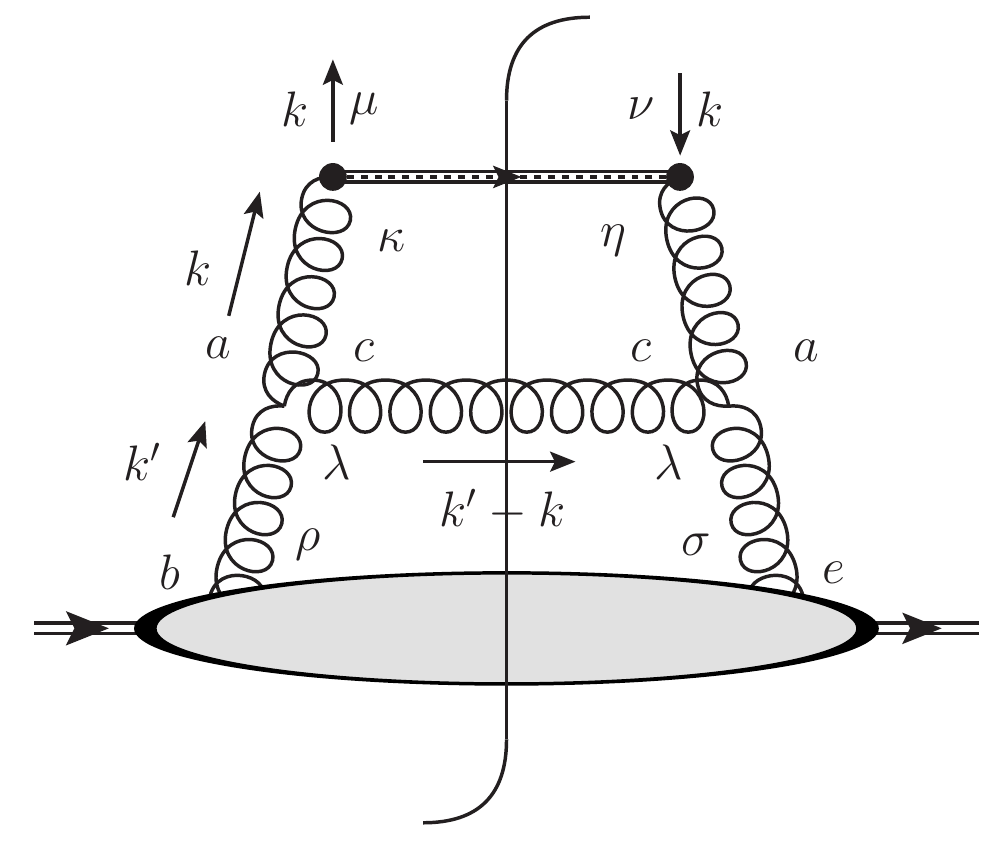} 
\includegraphics[width=4.5cm]{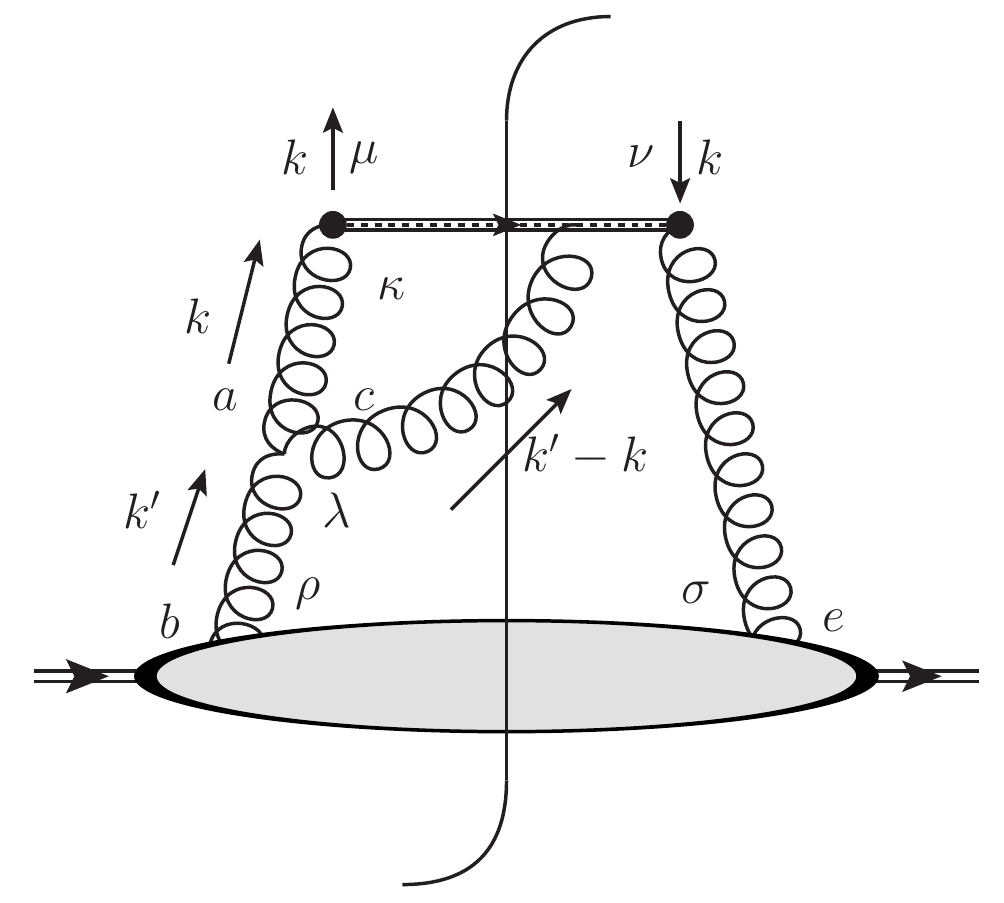}
\includegraphics[width=4.5cm]{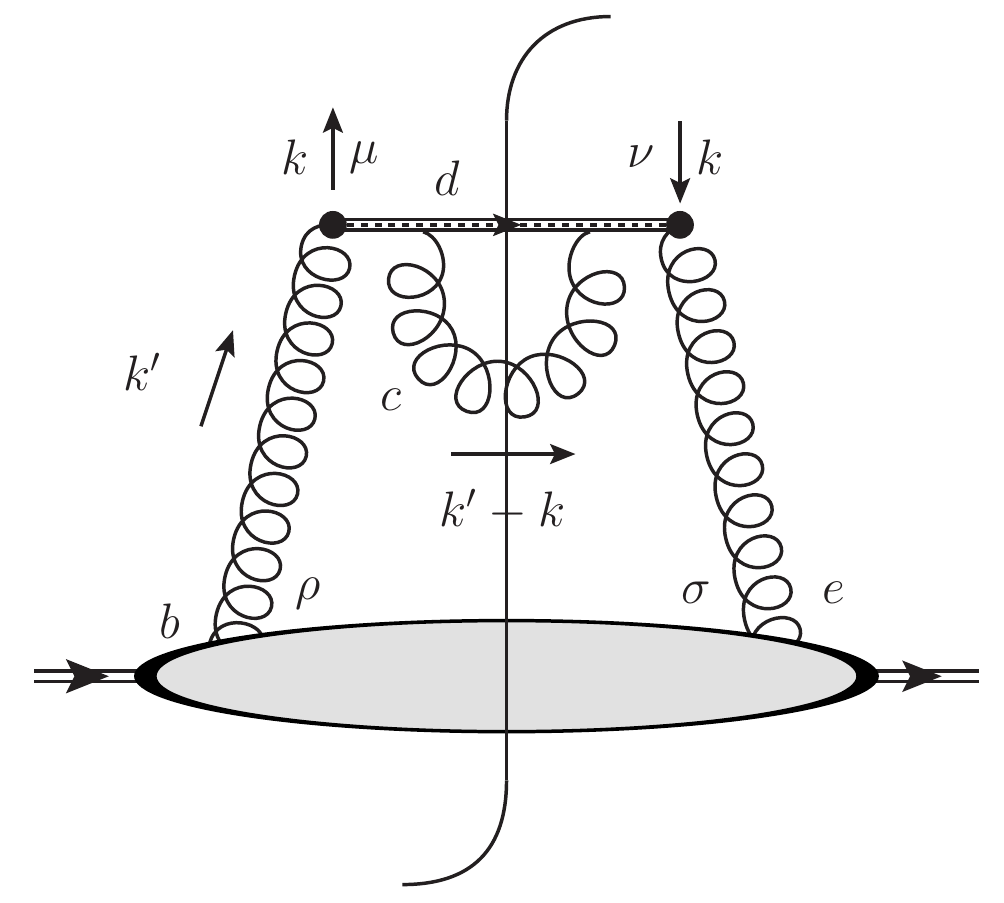}
\caption{Feynman diagrams contributing to the tail of the gluon correlator.}\label{fig:diagrams}
\end{figure}

The degree of polarization is defined as the linearly polarized gluon distribution 
with respect to its upper bound \cite{Mulders:2000sh}, i.e.,
\begin{equation}
 \mc{P}(x,\kT^2,\zeta,\mu) \equiv 
 \frac{h_1^{\perp g}(x,\kT^2,\zeta,\mu)}{\frac{2M_p^2}{\kT^2} f_1^g(x,\kT^2,\zeta,\mu)},
 \text{ such that } |\mc{P}|\leq 1. 
\end{equation}
The tail of the gluon correlator ($\kT^2 \gg M_p^2$)
can be expressed in terms of the Feynman diagrams in Fig.\ \ref{fig:diagrams} as
\begin{equation}
  \Phi_\text{tail}^{\mu\nu}(x,\kT,\zeta,\mu) = \frac{\int\! \d (k\cdot P)}{(k\cdot n)^2} 
  \left( \text{diag 1} + \text{diag 2} + \text{diag 3}\right).
\end{equation}
The contributions from the three diagrams to the tail are
\begin{align}
 \Phi^{\mu\nu}_{\mathrm{diag 1}} =& \frac{4 \pi \alpha_s}{(k{\cdot} n)^2} f^{abc}f^{aec}\!\!
 \int\frac{\d^4(k^\prime-k) \d(k{\cdot} P)}{(2\pi)^3} \delta \Big[ (k^\prime-k)^2 \Big]  
 \frac{\Big[k\cdot n g^\mu_\kappa - k^\mu n_\kappa\Big]
 \Big[k\cdot n g^\nu_\eta - k^\nu n_\eta\Big]}{k^4} \Phi_{\rho\sigma}^{be} (k^\prime)\nonumber\\
 \Big[g^{\kappa\lambda} (k^\prime& -2k)^\rho + g^{\lambda\rho}(k-2k^\prime)^\kappa + g^{\rho\kappa}(k^\prime + k)^\lambda\Big]
 \Big[g_\lambda^\eta (2k-k^\prime)^\sigma + g_\lambda^\sigma (2k^\prime-k)^\eta + g^{\eta\sigma}(-k^\prime-k)_\lambda\Big]
\nonumber\\
 \Phi^{\mu\nu}_{\mathrm{diag 2}} =& \frac{4 \pi \alpha_s}{(k{\cdot} n)^2}  f^{abc}f^{ace}\!\!
 \int\frac{\d^4(k^\prime-k) \d(k{\cdot} P)}{(2\pi)^3} \delta \Big[ (k^\prime-k)^2 \Big]  
 \Big[k\cdot n g^\mu_\kappa - k^\mu n_\kappa\Big] \Big[k\cdot n g^{\nu\sigma} - k^{\prime\nu} n^\sigma\Big]\Phi_{\rho\sigma}^{be} (k^\prime)\nonumber\\
 &\frac{1}{k^2} \frac{n_\lambda}{(k^\prime-k)\cdot n}
 \Big[g^{\kappa\lambda} (k^\prime -2k)^\rho + g^{\lambda\rho}(k-2k^\prime)^\kappa + g^{\rho\kappa}(k^\prime + k)^\lambda\Big]
 + \left(\mu\leftrightarrow\nu\right)
\nonumber\\
 \Phi^{\mu\nu}_{\mathrm{diag 3}} =& \frac{4 \pi \alpha_s}{(k{\cdot} n)^2}  f^{bcd}f^{ced}\!\!
 \int\frac{\d^4(k^\prime-k) \d(k{\cdot} P)}{(2\pi)^3} \delta \Big[ (k^\prime-k)^2 \Big]  
 \Big[k\cdot n g^{\mu\rho} - k^{\prime\mu} n^\rho \Big] \Big[k\cdot n g^{\nu\sigma} - k^{\prime^\nu} n^\sigma\Big]
 \nonumber\\
 &\Phi_{\rho\sigma}^{be} (k^\prime)
 \frac{n^2}{[(k^\prime-k)\cdot n]^2}
 \end{align}
In the following calculation, the momenta $k$, $k^\prime$ and $n$ will be parameterized as
\begin{align}\label{eq:kkpparam}
 k 		&= x P + k_\sT + \frac{k\cdot P}{P\cdot P^\prime} P^\prime, \nonumber\\
 k^\prime 	&= y P +k^\prime_\sT + \frac{k^\prime \cdot P}{P\cdot P^\prime} P^\prime, \nonumber\\
 n		&= \frac{\sqrt{n^2}}{\zeta} P + \frac{\sqrt{n^2}\zeta}{2 P\cdot P^\prime} P^\prime,
\end{align}
in terms of the light-like vectors $P$ and $P^\prime$. 
Note that $k_\sT^2 = - \kT^2$.
In the parameterization of the Lorentz structure we do not want $P$ and $P^\prime$,
but rather $k_\sT$, $k$ and $n$, so the inverse relations are also useful,
\begin{align}
 P &= - \frac{\zeta^2}{2 k\cdot P - x \zeta^2} (k-k_\sT) + \frac{2\zeta P\cdot P^\prime}{2 k\cdot P - x \zeta^2} \frac{n}{\sqrt{n^2}},\nonumber\\
 P^\prime &= \frac{2 P\cdot P^\prime}{2 k\cdot P - x \zeta^2} (k-k_\sT) - \frac{2x\zeta P\cdot P^\prime}{2 k\cdot P - x \zeta^2}\frac{n}{\sqrt{n^2}}.
\end{align}
The phase space becomes in this parameterization
\begin{equation}
 \int\d^4 (k^\prime-k)	
		= \frac{1}{P\cdot P^\prime} \int \d(k^\prime\cdot P)\, \d(k^\prime\cdot P^\prime)\, \d^2 k_\sT^\prime
		= \int \d(k^\prime\cdot P)\, \d y\, \d^2 k_\sT^{\prime}
\end{equation}
The delta function that sets the emitted gluon on-shell will be removed by the
integration over $k\cdot P$, i.e.,
\begin{equation}
 \int \d(k\cdot P) \delta \Big[ (k^\prime -k)^2 \Big]
 = \int \d(k\cdot P) \delta \Big( 2(y-x)(k^\prime -k)\cdot P + (k_\sT^\prime - k_\sT)^2 \Big)
 = \frac{1}{2(y-x)},
\end{equation}
which sets
\begin{equation}\label{eq:kdotP}
 k\cdot P = \frac{(k_\sT^\prime - k_\sT)^2}{2(y-x)} + k^\prime\cdot P.
\end{equation}

The next step is to use the fact that $k^\prime\cdot P \ll k_\sT^2$ and $k_\sT^{\prime 2} \ll k_\sT^2$.
We will make a zeroth order expansion in these small variables of the integrand, such that
it does not depend on them anymore.
This allows us to move everything out of the integral and express the tail of the TMD correlator 
in terms of the collinear correlator,
\begin{equation}
 \int\!\d^2 k_\sT^\prime \int\! \d(k^\prime\cdot P)\, \Phi^{\rho\sigma} (k^\prime) 
    = \Phi^{\rho\sigma}(y) \nonumber 
    = -\frac{1}{2y} g_\sT^{\rho\sigma} f_1^g(y,\mu_0),
\end{equation}
where $f_1^g(y,\mu_0)$ is the collinear gluon parton distribution function
evaluated at a scale $\mu_0^2 \ll k_\sT^2$.
The resulting expressions can be cast in the following form
\begin{multline}\label{eq:correxp}
 \Phi^{\mu\nu}_{\textrm{diag } i}(x,\kT,\zeta,\mu) = \frac{\alpha_s C_A }{4\pi^2}
 \int\!\d y \Big [
 A_i\, g_\sT^{\mu\nu} + B_i\, k_\sT^\mu k_\sT^\nu + C_i (n\cdot k n^\mu -n^2 k^\mu) (n\cdot k n^\nu -n^2 k^\nu)\\
 + D_i \big[ (n\cdot k n^\mu -n^2 k^\mu)k_\sT^\nu + k_\sT^\nu (n\cdot k n^\mu -n^2 k^\nu)\big]
 \Big ] f_1^g(y,\mu_0),
 \end{multline}
in which the following coefficients are nonzero,
\begin{align}
  A_1 &= \frac{3y-2x}{k_\sT^2 y^3}, &
  B_1 &= \frac{4 (y-x)^3(\zeta^4 y^2 - 2\zeta^2 k_\sT^2)}{k_\sT^4 y^3 [k_\sT^2 + \zeta^2 x (x-y)]^2},\nonumber\\
  C_1 &= \frac{16 \zeta^4 (y-x)^3 (2x^2 - 2xy + y^2)}{n^4 y^3 [k_\sT^4 - \zeta^4 x^2 (x-y)^2]^2}, &
  D_1 &= -\frac{8\zeta^2 (x-y)^3 (\zeta^2 (x^2-xy+y^2) - k_\sT^2)}{k_\sT^2 n^2 y^3 [k_\sT^2 - \zeta^2 x(y-x)]^2
	[k_\sT^2 + \zeta^2 x(y-x)]},\nonumber\\
  A_2 &= \frac{k_\sT^2 + \zeta^2 (y^2 - x^2)}{k_\sT^2 y^2 [\zeta^2 (x-y)^2 - k_\sT^2]}, &
  D_2 &= - \frac{4\zeta^2 (x-y)^2}{k_\sT^2 n^2 y^2 [\zeta^2 (x-y)^2 - k_\sT^2][k_\sT^2 + \zeta^2 x (y-x)]},\nonumber\\
  A_3 &= \frac{2\zeta^2 (y-x)}{y[k_\sT^2 - \zeta^2 (x-y)^2]^2}.
\end{align}
Comparing Eq.\ \eqref{eq:TMDcorrelator} with Eq.\ \eqref{eq:correxp} one can read off that
the distribution functions are, in terms of these coefficients, given by
\begin{align}
 f_{1 \textrm{tail}}^g(x,\kT^2,\zeta,\mu) &= - \frac{x \alpha_s C_A}{2 \pi^2} \int\! \d y\, 
 \sum_i \left(A_i + \frac{k_\sT^2}{2} B_i\right) f_1^g(y,\mu_0), \nonumber\\
 h_{1 \textrm{tail}}^{\perp g}(x,\kT^2,\zeta,\mu) &=  \frac{x \alpha_s C_A M_p^2}{2 \pi^2} \int\! \d y\, 
 \sum_i B_i f_1^g(y,\mu_0).
\end{align}
Note that at this order the tail of the distribution functions does not depend on
the UV regulator $\mu$.

\begin{figure}
\centering
\includegraphics[width=4.5cm]{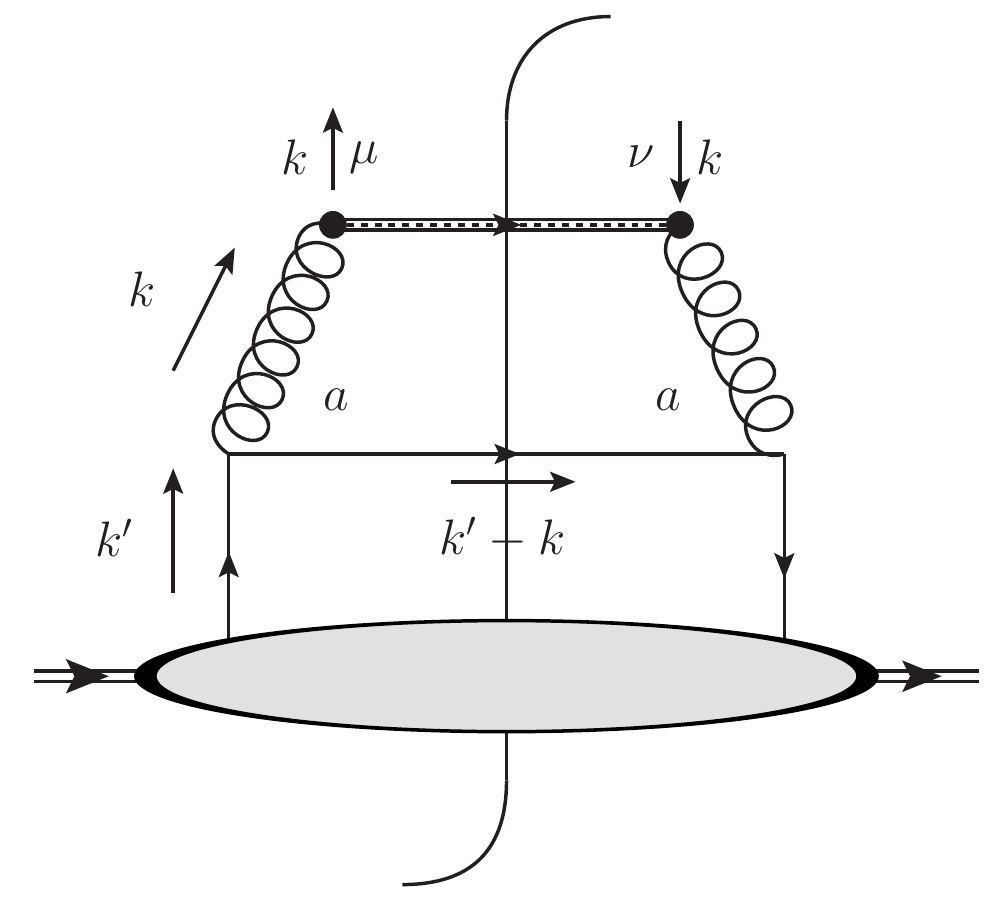} 
\includegraphics[width=4.5cm]{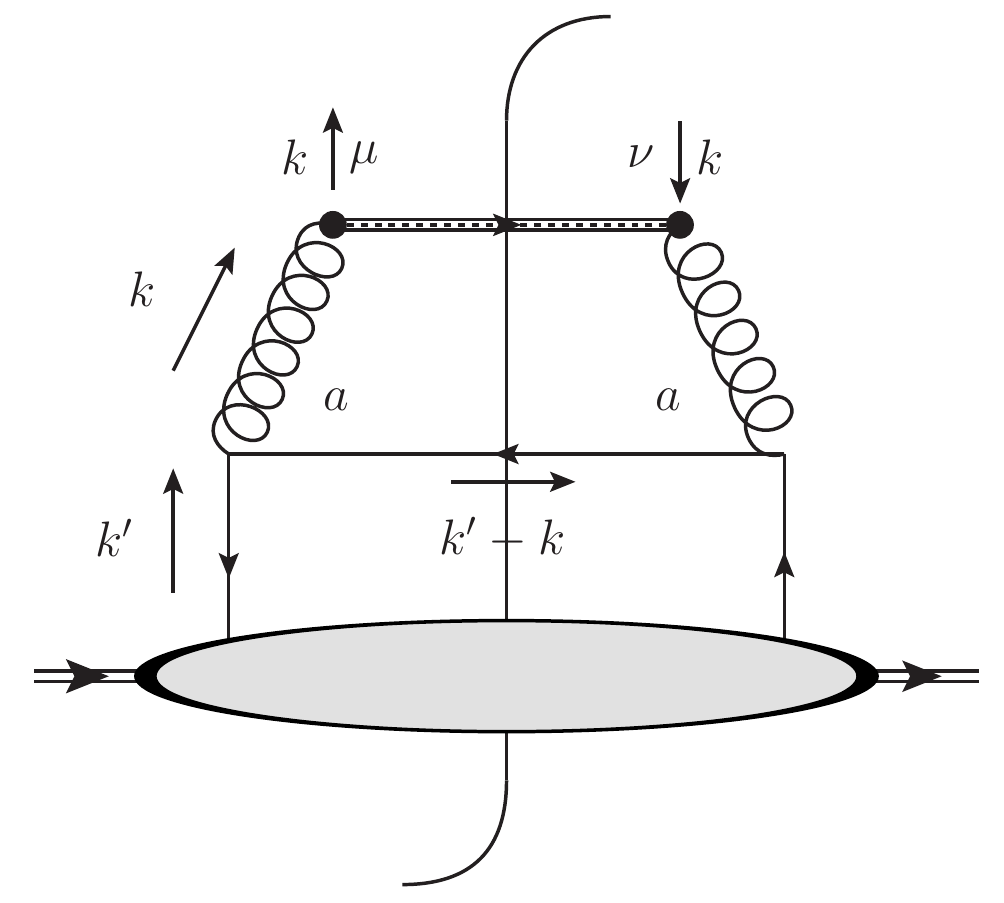}
\caption{Quark initiated Feynman diagrams contributing to 
the tail of the gluon correlator.}\label{fig:quarkdiagrams}
\end{figure}

\subsection*{Quark contribution}

There will also be a contribution from the quark and anti-quark initiated diagrams
depicted in Fig.\ \ref{fig:quarkdiagrams}. 
Those contributions can be calculated in the same way, yielding
\begin{multline}\label{eq:correxpquark}
 \Phi^{\mu\nu}_{\mathrm{quark}}(x,\kT) = \frac{ \alpha_s C_F }{4 \pi^2}
 \int\!\d y \Big [
 A_q g_\sT^{\mu\nu} + B_q k_\sT^\mu k_\sT^\nu + C_q (n\cdot k n^\mu -n^2 k^\mu) (n\cdot k n^\nu -n^2 k^\nu)\\
 + D_q \big[ (n\cdot k n^\mu -n^2 k^\mu)k_\sT^\nu + k_\sT^\nu (n\cdot k n^\mu -n^2 k^\nu)\big]
 \Big ] \sum_{q,\bar{q}} f_1^q(y,\mu_0),
 \end{multline}
with the coefficients
\begin{align}
 A_q &= \frac{1}{k_\sT^2 y^2}, &
 B_q &= -\frac{4 \zeta^2 (x-y)^2 [k_\sT^2 + \zeta^2 y (x-y)]}{k_\sT^4 y^2 [k_\sT^2 + \zeta^2 x (x-y)]^2},\nonumber\\
 C_q &= - \frac{16 \zeta^4 (x-y)^5}{n^4 y^2 [k_\sT^4 - \zeta^4 x^2 (x-y)^2]^2}, &
 D_q &= \frac{4 \zeta^2 (x-y)^3 [k_\sT^2 - \zeta^2 (x-2y)(x-y)]}{k_\sT^2 n^2 y^2 [k_\sT^2 + \zeta^2 x (x-y)]^2
	[k_\sT^2 + \zeta^2 x (y-x)]}.
\end{align}
Their contribution to the distribution functions is given by
\begin{align}
f_{1 \text{tail}}^g(x,\kT^2,\zeta,\mu) 		
 &= - \frac{x \alpha_s C_F}{2 \pi^2} \int\! \d y\, 
  \left(A_q + \frac{k_\sT^2}{2} B_q\right) \sum_{q,\bar{q}}\, f_1^q(y,\mu_0), \nonumber\\
h_{1 \text{tail}}^{\perp g}(x,\kT^2,\zeta,\mu) 	
 &= \frac{x \alpha_s C_F M_p^2}{2 \pi^2} \int\! \d y\, B_q\, \sum_{q,\bar{q}}\, f_1^q(y,\mu_0).
\end{align}

\subsection*{The $\zeta\to\infty$ limit}

Calculations in the literature are often performed in the 
$\zeta\to\infty$ limit, in which we get
\begin{align}
 f_{1 \text{tail}}^g(x,\kT^2,\zeta,\mu)  
 =& - \frac{\alpha_s}{\pi^2 k_\sT^2} 
	\int_x^1\!  \frac{\d z}{z}\Bigg[
	C_A \bigg\{ \frac{1-z}{z} + z(1-z) + \frac{z}{(1-z)^+}
	- \frac{1}{2}\delta(z-1)\times \nonumber\\
	&\left( 1 + \log\left[\frac{-k_\sT^2}{x^2 (1-x)^2 \zeta^2}\right] \right) \bigg\}
	f_1^g \left(\frac{x}{z},\mu_0\right)
	+ C_F \frac{1+(1-z)^2}{2z} \sum_{q,\bar{q}} 
	f_1^q\left(\frac{x}{z},\mu_0\right)\Bigg],
	\nonumber\\
 h_{1 \text{tail}}^{\perp g}(x,\kT^2,\zeta,\mu) 
 =& \frac{\alpha_s}{\pi^2 k_\sT^2} \frac{2 M_p^2}{k_\sT^2}
		      \int_x^1\!  \frac{\d z}{z} \frac{1-z}{z} 
		      \Bigg[ C_A 
		      f_1^g \left(\frac{x}{z},\mu_0\right)
		      + C_F \sum_{q,\bar{q}} 
		      f_1^q\left(\frac{x}{z},\mu_0\right) \Bigg].
\end{align}

\section*{Results}

\begin{figure}
\centering
\includegraphics[height=4.4cm]{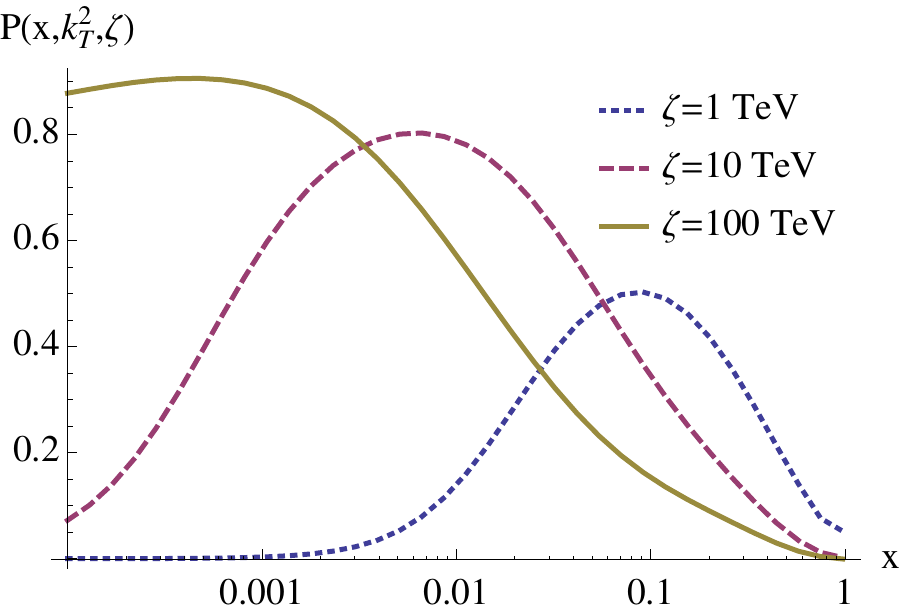} 
\includegraphics[height=4.4cm]{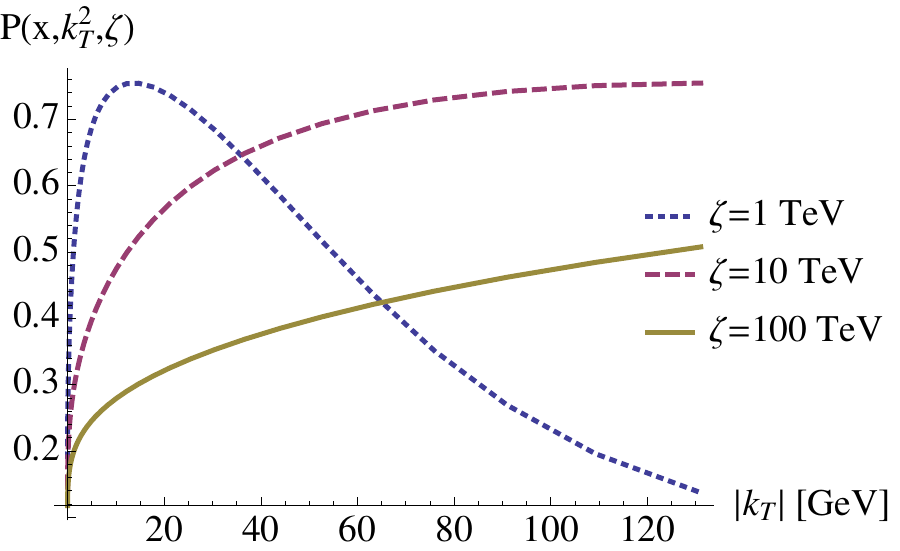}
\caption{Degree of polarization plotted as a function of $x$ for fixed $|\kT|=100$ GeV (left)
and as function of $|\kT|$ at fixed $x=125/8000$ (right) for various choices of $\zeta$.}\label{fig:dop}
\end{figure}

Adding the gluon and (anti)quark initiated contributions,
the degree of polarization can, for arbitrary $\zeta$ and large $\kT$, be written as
\begin{equation}
  \mathcal{P}(x,\kT^2,\zeta,\mu)
    = \frac{\frac{k_\sT^2}{2} \int_x^1\! \d y \left[ C_A B_1 f_1^g(y,\mu_0) + C_F B_q \sum\limits_{q,\bar{q}} f_1^q(y,\mu_0)\right]}
    {\int_x^1\! \d y\left[ C_A \left( A_1 + A_2 + A_3 + \frac{k_\sT^2}{2} B_1\right) f_1^g(y,\mu_0)
    + C_F \left( A_q + \frac{k_\sT^2}{2} B_q\right) \sum\limits_{q,\bar{q}} f_1^q(y,\mu_0)\right]},
 \end{equation}
up to corrections of $\mc{O}(\alpha_s)$.
The degree of polarization is plotted in Fig.\ \ref{fig:dop} as a function of $x$ for fixed $|\kT|=100$ GeV
and as function of $|\kT|$ at fixed $x=125/8000$ for various choices of $\zeta$,
calculated from the MSTW 2008 parton distributions \cite{Martin:2009iq} 
evaluated at $\mu_0 = 2$ GeV.
One can see that at the values of $x$ and $\zeta$ relevant for Higgs production at the LHC 
($\sim 125/8000$ and $\sim 12$ TeV) the degree of polarization is substantial,
reaching up to values larger than 70\%.

\section*{Summary}

Gluons inside an unpolarized proton are in general linearly polarized with a direction
and magnitude depending on their transverse momentum. 
This polarization can, in principle, be used to determine the spin and parity of 
the newly found Higgs-like boson at the LHC as proposed in \cite{Boer:2013fca}.
We have here elaborated on the perturbative calculation of the degree of polarization,
which was used in \cite{Boer:2013fca}.
Numerical results have been given for the degree of polarization as a function of 
the longitudinal momentum fraction $x$ and the transverse momentum $|\kT|$, 
for various choices of the gauge link direction (or hadron energy) $\zeta^2 \equiv (2n{\cdot}P)^2/n^2$.
At the values of $x$ and $\zeta$ relevant for Higgs production at the LHC 
the degree of polarization is substantial, reaching up to values larger than 70\%.

\begin{acknowledgments}
This work was supported in part by the German Bundesministerium f\"{u}r Bildung und For\-schung (BMBF),
grant no. 05P12VTCTG.
\end{acknowledgments}

\end{document}